\documentclass[
    ,final            
  ]
  {aipproc}

\layoutstyle{6x9}

\usepackage{amssymb}

\begin{document}

\title{FlashCam: A fully digital camera for CTA telescopes}




\author{G. P{\"u}hlhofer}{
  address={Institut f{\"u}r Astronomie und Astrophysik, Kepler Center for Astro and Particle Physics, Eberhard-Karls-Universit{\"a}t, Sand 1, D 72076 T{\"u}bingen, Germany}
}
\author{C. Bauer}{
  address={Max-Planck-Institut f\"ur Kernphysik, P.O. Box 103980, D 69029 Heidelberg, Germany}
}
\author{A. Biland}{
  address={ETH Z{\"u}rich, Inst. for Particle Physics, Schafmattstr. 20, CH-8093 Z{\"u}rich, Switzerland}
}
\author{D. Florin}{
  address={Physik-Institut, Universit{\"a}t Z{\"u}rich, Winterthurerstrasse 190, 8057 Z{\"u}rich, Switzerland}
}
\author{C. F{\"o}hr}{
  address={Max-Planck-Institut f\"ur Kernphysik, P.O. Box 103980, D 69029 Heidelberg, Germany}
}
\author{A. Gadola}{
  address={Physik-Institut, Universit{\"a}t Z{\"u}rich, Winterthurerstrasse 190, 8057 Z{\"u}rich, Switzerland}
}
\author{G. Hermann}{
  address={Max-Planck-Institut f\"ur Kernphysik, P.O. Box 103980, D 69029 Heidelberg, Germany}
}
\author{C. Kalkuhl}{
  address={Institut f{\"u}r Astronomie und Astrophysik, Kepler Center for Astro and Particle Physics, Eberhard-Karls-Universit{\"a}t, Sand 1, D 72076 T{\"u}bingen, Germany}
}
\author{J. Kasperek}{
  address={Faculty of Electrical Engineering, Automatics, Computer Science and Electronics, AGH University of Science and 
Technology, Al. Mickiewicza 30, 30-059 Cracow, Poland}
}
\author{T. Kihm}{
  address={Max-Planck-Institut f\"ur Kernphysik, P.O. Box 103980, D 69029 Heidelberg, Germany}
}
\author{J. Koziol}{
  address={Jagiellonian University, ul. Orla 171, 30-244 Cracow, Poland}
}
\author{A. Manalaysay}{
  address={Physik-Institut, Universit{\"a}t Z{\"u}rich, Winterthurerstrasse 190, 8057 Z{\"u}rich, Switzerland}
}
\author{A. Marszalek}{
  address={Jagiellonian University, ul. Orla 171, 30-244 Cracow, Poland}
}
\author{P.J. Rajda}{
  address={Faculty of Electrical Engineering, Automatics, Computer Science and Electronics, AGH University of Science and 
Technology, Al. Mickiewicza 30, 30-059 Cracow, Poland}
}
\author{T. Schanz}{
  address={Institut f{\"u}r Astronomie und Astrophysik, Kepler Center for Astro and Particle Physics, Eberhard-Karls-Universit{\"a}t, Sand 1, D 72076 T{\"u}bingen, Germany}
}
\author{S. Steiner}{
  address={Physik-Institut, Universit{\"a}t Z{\"u}rich, Winterthurerstrasse 190, 8057 Z{\"u}rich, Switzerland}
}
\author{U. Straumann}{
  address={Physik-Institut, Universit{\"a}t Z{\"u}rich, Winterthurerstrasse 190, 8057 Z{\"u}rich, Switzerland}
}
\author{C. Tenzer}{
  address={Institut f{\"u}r Astronomie und Astrophysik, Kepler Center for Astro and Particle Physics, Eberhard-Karls-Universit{\"a}t, Sand 1, D 72076 T{\"u}bingen, Germany}
}
\author{P. Vogler}{
  address={ETH Z{\"u}rich, Inst. for Particle Physics, Schafmattstr. 20, CH-8093 Z{\"u}rich, Switzerland}
}
\author{A. Vollhardt}{
  address={Physik-Institut, Universit{\"a}t Z{\"u}rich, Winterthurerstrasse 190, 8057 Z{\"u}rich, Switzerland}
}
\author{Q. Weitzel}{
  address={ETH Z{\"u}rich, Inst. for Particle Physics, Schafmattstr. 20, CH-8093 Z{\"u}rich, Switzerland}
}
\author{K. Winiarski}{
  address={Faculty of Electrical Engineering, Automatics, Computer Science and Electronics, AGH University of Science and 
Technology, Al. Mickiewicza 30, 30-059 Cracow, Poland}
}
\author{K. Zietara}{
  address={Jagiellonian University, ul. Orla 171, 30-244 Cracow, Poland}
}
\author{the CTA consortium}{
  address={See http://www.cta-observatory.org/?q=node/342 for full author \& affiliation list}
}

\begin{abstract}
The future Cherenkov Telescope Array (CTA) will consist of several tens of telescopes of different mirror sizes. CTA will provide next generation sensitivity to very high energy photons from few tens of GeV to $>$100 TeV. Several focal plane instrumentation options are currently being evaluated inside the CTA consortium. In this paper, the current status of the FlashCam prototyping project is described. FlashCam is based on a fully digital camera readout concept and features a clean separation between photon detector plane and signal digitization/triggering electronics.

\end{abstract}

\maketitle


\section{The FlashCam concept and camera architecture}

The FlashCam approach is unique
amongst the camera projects pursued inside CTA,
because signal processing inside the camera is fully digital. Signal digitization and trigger processing are jointly performed in one single readout chain per camera pixel. For a group of pixels, such a chain consists of Flash ADCs and a Field-Programmable Gate Array (FPGA) module, both commercially available at low cost. A camera-wide event trigger is subsequently computed from the digitized signals.
The camera data are transferred from the camera front-end to a camera server using a
high-performance Ethernet network protocol. Such a fully digital approach using state-of-the-art components provides accurate and flexible triggering and an easily scalable architecture in a cost-effective way. It also permits to cover the full dynamic range per pixel with only one readout channel, respectively, using compression of high amplitude signals in the preamplifier and recovery of the charge through signal processing in the FPGA.

The architecture of FlashCam is based on the concept of a {\em horizontal} integration of its components. Here, the photon detector plane (PDP) is a self-contained unit, consisting of photon detectors, high voltage supplies, preamplifiers, and control logic, which are sustained by a monolithic mechanical carrier. 
The signal digitization and triggering electronics are organized in boards and mini-crates, which are separated from the photon detector plane. 
Such an approach allows adaption of various photon detectors and pixel pitches and minimizes the amount of electronics directly at the focal plane, thereby reducing weight and cooling issues there, and also results in cost savings compared to other approaches.
The electronics crates are kept either at the rear side of the camera body (in case of a photo-multiplier-(PMT)-based PDP), or could be completely detached from a very compact focal-plane instrument that is based e.g. on Si-PMTs.

In Fig.\,\ref{fig:moboplusfadc_camerabody}, right, a preliminary mechanical concept
for a medium-sized telescope (MST) 1764 PMT pixel camera is shown.
The PDP is passively cooled, while 
crates are cooled by air ventilation. Options for heat exchange to the camera exterior -- assuming a near-sealed camera -- are being evaluated.
Development work towards a Si-PMT-based camera using the FlashCam concept and electronics is being pursued in parallel.

\section{Signal digitization, trigger, and data transmission}

The entire FlashCam electronics is based on commercially available microchips, in essence FADCs for signal digitization and FPGAs for further signal and trigger processing. The key point of the concept is to perform all pixel signal processing purely with digitized information. In particular, the trigger decision is computed solely from the digitized signals.
For such a camera, low power ($<$0.5\,W/channel) 12bit FADCs are currently available with sampling speeds of up to $\sim$250\,MHz. Through extensive simulations (including realistic time jitter and night sky background conditions) it could be shown that trigger performance as well as the pixel amplitude and timing resolution of such a system is fully competitive with higher (e.g. $1-2$\,GHz) sampling speed systems.

\begin{figure}
  \label{fig:moboplusfadc_camerabody}
  \resizebox{!}{8.5pc}{\includegraphics{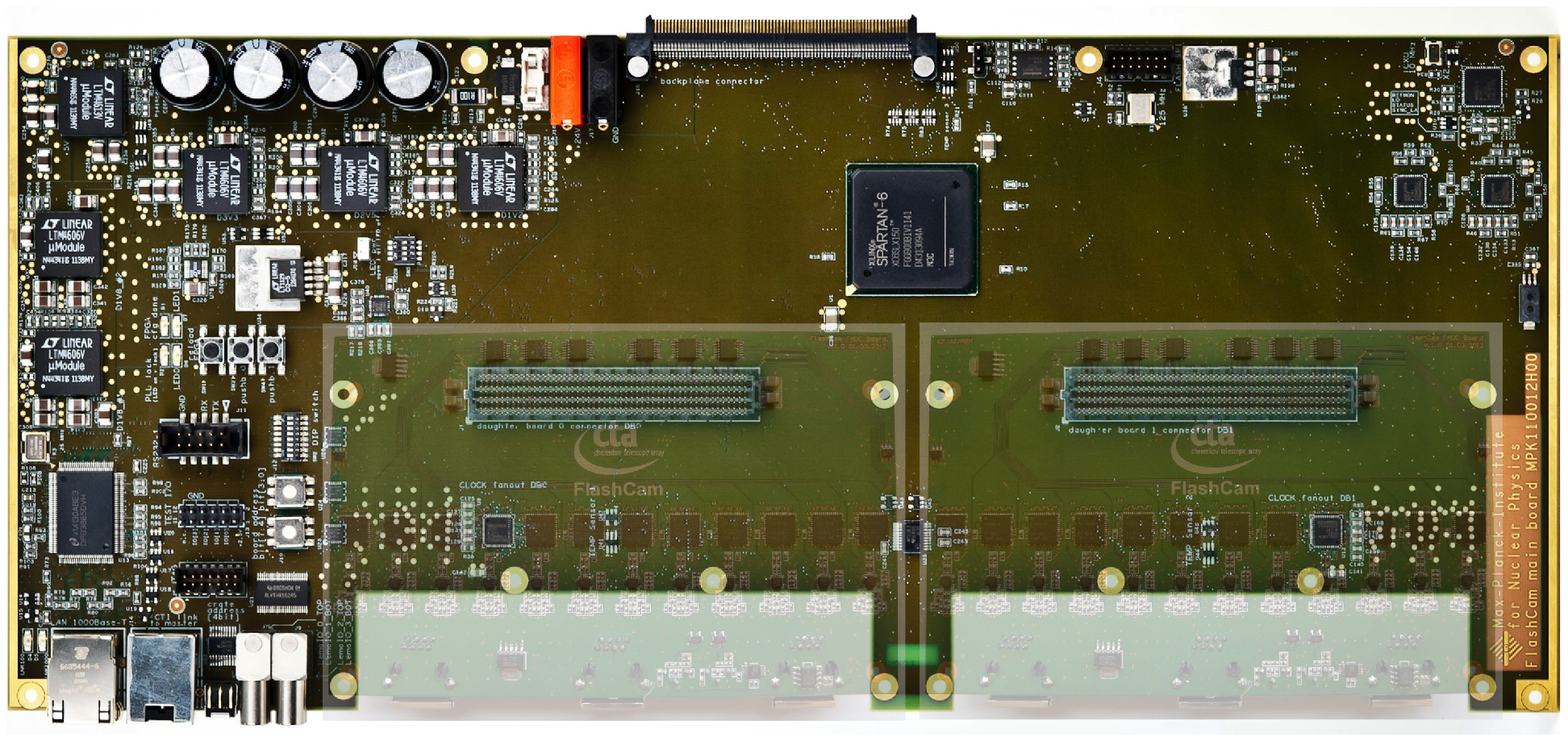}}
  \hspace{0.3pc}
  \resizebox{!}{12pc}{\raisebox{-4.5pc}{\includegraphics{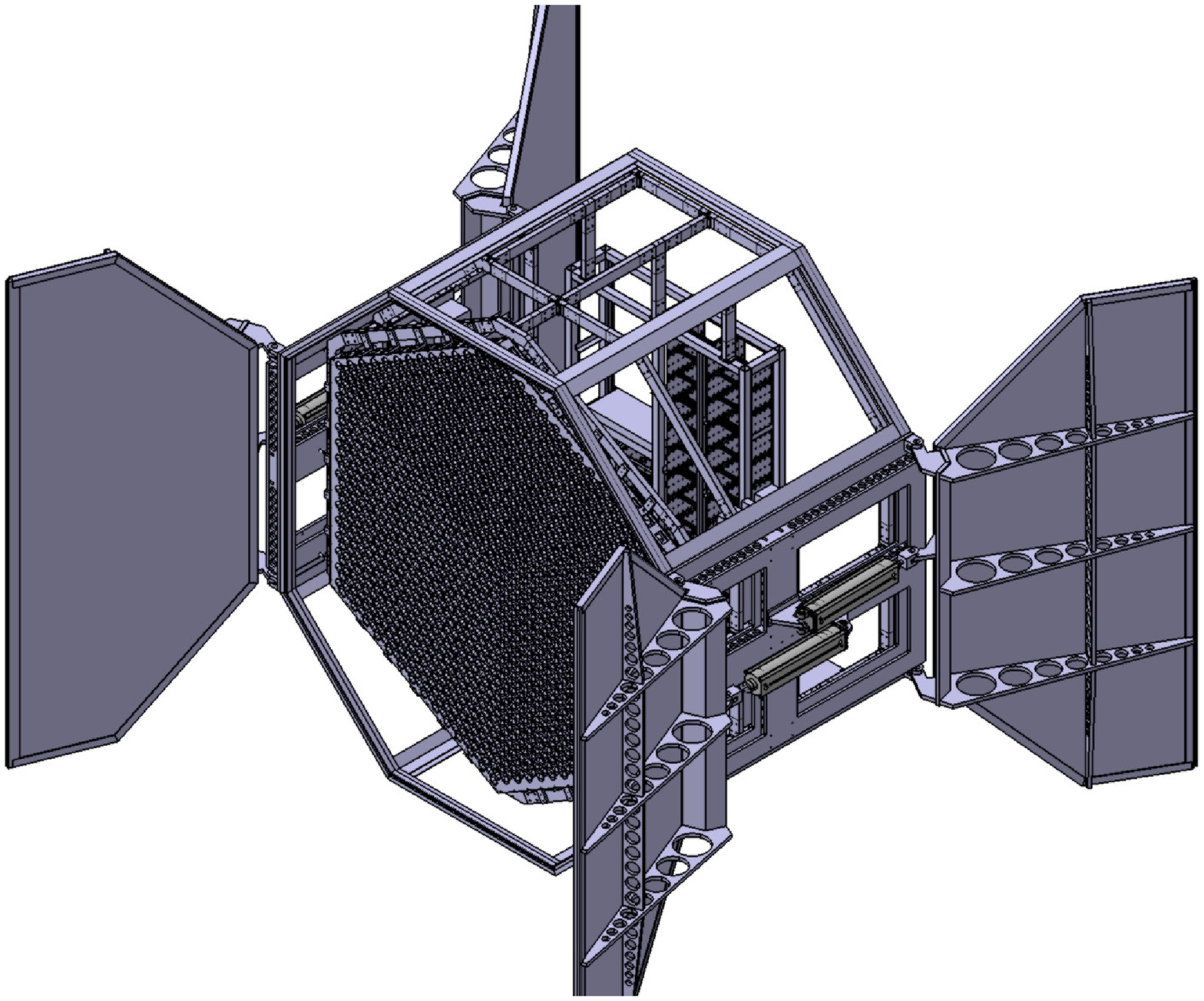}}}
\caption{
{\bf Left:} FlashCam motherboard (14 layers PCB stack), with one backplane connector (32 1\,Gb/s LVDS pairs) and two daughter board connectors (99 1\,Gb/s LVDS pairs each). Two 12 channel FADC daughter boards (10 layers PCB stack), with 3 RJ45 connectors for analogue signals and a 400 pin motherboard connector, can be mounted on the motherboard (semi-transparent insets).
{\bf Right:} Drawing of a possible implementation of a FlashCam-based MST camera. Front and rear doors are open, isolation panels as well as light guides are not shown. A human can enter the camera body and can access the rear of the photon detector plane as well as all electronic crates, for installation and in case of repair work.
\vspace{-1.5pc}
}
\end{figure}

The FlashCam readout and trigger electronics is based on the development of one motherboard with a Spartan-6 FPGA, 
which carries either FADC digitizer, trigger interface, or clock/master trigger mezzanine (``daughter'') boards (Fig.\,\ref{fig:moboplusfadc_camerabody}, left).
The expected data rate of such a system ($\sim$700\,MB/s for a
1764 pixel camera and a trigger and readout rate of 10\,kHz) allows the full pixel event information to be transmitted over standard gigabit Ethernet infrastructure without any data reduction. The up to 96 FADC boards have a 1\,Gbit Ethernet connection each, which will be routed through one or two 10\,Gbit Ethernet lines for data transmission to the camera server. The transmission protocol is raw Ethernet to avoid any overhead of higher level protocols. 
Current tests of the data transmission routinely operate at 
rates of up to 1.8 GByte/s, well beyond requirements.

\section{PMT-based photon detector plane modules}

\begin{figure}[t]
  \label{fig:pdpmodule_amplituderesolution}
  \resizebox{!}{13pc}{\reflectbox{\includegraphics{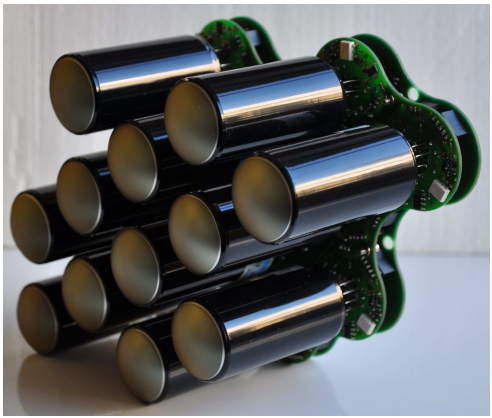}}}
   \resizebox{!}{14pc}{\raisebox{-3.5pc}{\includegraphics{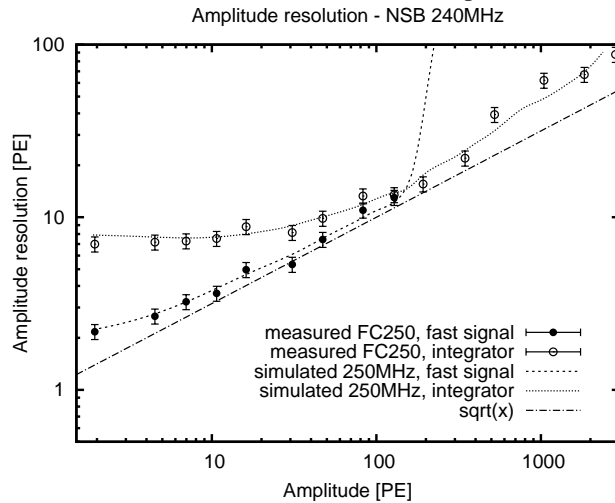}}}
\caption{
{\bf Left:} Front view of the FlashCam 12-pixel PDP module. PMTs are from Hamamatsu (R11920, the current baseline option for CTA).
{\bf Right:} Simulated (lines) and measured (data points) pixel amplitude resolution. Simulations and measurements are for 240\,MHz night sky background (NSB), for the CTA PMT, and using a search window of $\pm1$\,ns and an integration window of 8\,ns. Air shower light was emulated with a fast (FWHM$<$300\,ps) laser and a filter wheel, 
and NSB
with a light bulb.
\vspace{-1pc}
}
\end{figure}

The FlashCam PDP concept for PMTs is based on mechanical modules of 12 pixels (Fig.\ref{fig:pdpmodule_amplituderesolution}, left). Each module contains PMTs, preamplifiers, HV supply, and a micro-controller for slow control and monitoring. Each module weighs about 1.1\,kg and dissipates only 2.3\,W. 
Two such PDP modules are served by one FADC electronics board. The analogue signals of the PMTs are transmitted via CAT 5/6 cables (four differential lines per cable). Through one additional (9-pin D-sub) connector, CAN-bus control and 24\,V power supply are provided to each PDP module. Low ($\pm5$\,V) and high voltage are generated on the module, thereby avoiding expensive HV-rated connectors.

The HV regulation features a separation into two bias supply lines ($-1500$\,V and $-500$\,V) and supplying the last three dynodes via an emitter follower PNP transistor circuit. Compared to conventional designs, power consumption is considerably reduced here. Pixels are individually controllable (HV off, HV $-700$\,V .. $-1500$\,V). 
A HV design with even lower cost and power consumption is currently being evaluated as well.

\section{FlashCam Pixel Performance}
The FlashCam concept foresees the use of preamplifiers with non-linear gain characteristics, instead of splitting the signal into two separate channels with different (low and high) gains. This allows cost to be reduced (the low gain channel can be saved), and permits full flexibility should future simulations result in a 
dynamic range requirement
below the currently anticipated 3000 photoelectrons (pe) per pixel.
The baseline design is to run an operational amplifier with a high input resistor (1-2\,k$\Omega$ instead of 50\,$\Omega$). High amplitude signals ($\gtrsim200$\,pe) saturate the output of the amplifier, but the input signal can be restored since the output signal broadens with a defined recovery time. The analysis of the digitized signal therefore runs in two regimes: Below 200 pe, the full time-resolved signal can be stored 
(with a window of $\sim$60-80\,ns), and digital filtering performed at server 
level allows the signal amplitude to be 
computed. 
Above 200\,pe, the signal is reconstructed directly in the FPGA from an integration over $\sim$200\,ns.

The system was verified under realistic conditions, using a FlashCam PDP module and a 
demonstrator board with 250\,MHz sampling speed.
Comparisons of the measured data with simulated performance expectations (e.g. Fig.\,\ref{fig:pdpmodule_amplituderesolution}, right) show that the system is fully understood. The achieved values are well within requirements for CTA cameras.

\section{Camera-wide Event Trigger and Performance}
In FlashCam,
the event trigger is derived from the digitized pixel signals processed in FPGAs. The digitized signals are sent from
the eight FADC boards via backplane to the trigger board 
in the same crate. Each trigger board has an interface of 226 differential signal pairs connected to the FADC boards. Each of these lines can receive up to 1\,Gbit/s of data. The total transmission capacity of 12 trigger boards is around 2.7\,Tbit/s.

The trigger algorithm is freely programmable by the FPGA firmware in each FADC and trigger board. Various trigger algorithms can be exploited and improved even during run-time of the experiment. Simulations show that a pre-grouping of pixel trigger into pixel ``patch'' information can be used to reduce bandwidth requirements (number of pins) without significant loss of sensitivity. A patch size of three pixels provides excellent spatial trigger homogeneity and sensitivity combined with high versatility and thus is used for the design of the MST-scale FlashCam prototype. 
The 3-pixel-patch information 
can contain up to 8 bit of information and allows to code multiplicity and/or summed amplitude triggers (clipped sum trigger). Simulations show that the performance of this bandwidth is almost indistinguishable from full-bandwidth trigger data transmission. The patch information is sent at each clock cycle (4\,ns) to the trigger cards. 
There, a trigger decision from 3, 7, or 19 of these patches is computed. The final patches have a roundish form and can be computed around each 3-pixel-patch individually.


\begin{theacknowledgments}
  We gratefully acknowledge support from the agencies and organisations listed on this page: http://www.cta-observatory.org/?q=node/22.
\end{theacknowledgments}



\bibliographystyle{aipproc}   

\end{document}